\documentclass[aps,floats]{revtex4}
\usepackage{amsmath,amssymb}
\usepackage{graphicx,epsfig}
\usepackage[greek,english]{babel}

\begin{document}
\bibliographystyle {plain}

\def\oppropto{\mathop{\propto}} 
\def\opsimeq{\mathop{\simeq}}
\def\opoverderline{\mathop{\overline}}
\def\operarrow{\mathop{\longrightarrow}}
\def\opsim{\mathop{\sim}} 
\def\opmin{\mathop{\min}} 
\def\opmax{\mathop{\max}} 

\def\fig#1#2{\includegraphics[height=#1]{#2}}
\def\figx#1#2{\includegraphics[width=#1]{#2}}


\title{ On the simplest scale invariant Tree-Tensor-States  \\
preserving the quantum symmetries of the antiferromagnetic XXZ chain} 


\author{ C\'ecile Monthus }
 \affiliation{Institut de Physique Th\'{e}orique, 
Universit\'e Paris Saclay, CNRS, CEA,
91191 Gif-sur-Yvette, France}

\begin{abstract}

For the line of critical antiferromagnetic XXZ chains with coupling $J>0$ and anisotropy $0<\Delta \leq 1$, we describe how the block-spin renormalization procedure preserving the $SU_q(2)$ symmetry introduced by Martin-Delgado and Sierra [Phys. Rev. Lett. 76, 1146 (1996)] can be reformulated as the translation-invariant scale-invariant Tree-Tensor-State of the smallest dimension that is compatible with the quantum symmetries of the model. The properties of this Tree-Tensor-State are studied in detail via the ground-state energy, the magnetizations and the staggered magnetizations, as well as the Shannon-Renyi entropies characterizing the multifractality of the components of the wave function.

\end{abstract}

\maketitle

\section{ Introduction }

Quantum information ideas \cite{nielsen}
have profoundly changed the perspective on many condensed matter problems (see the book \cite{qi} and references therein).
In particular, various real-space renormalization procedures have been reinterpreted from the point of view of Tensor Network States
(see the reviews \cite{ver,cirac,vidal_intro,phd-evenbly,mera-review,hauru,orus} and references therein) with the following output for the ground state of one-dimensional quantum spin chains : the Density-Matrix-RG introduced by S. White \cite{white} 
corresponds to Matrix-Products-States that are well adapted to describe non-critical states
displaying area-law entanglement, while the traditional block-spin renormalization for critical points corresponds to scale-invariant Tree-Tensor-States.
In addition, this way of thinking has produced completely new types of renormalization procedures, like the multi-scale-entanglement-renormalization-ansatz (MERA)
\cite{vidal_mera,evenbly_mera} for critical models, where 'disentanglers' between blocks are introduced besides the block-coarse-graining operators already present in Tree-Tensor-States (more details can be found in the reviews \cite{vidal_intro,phd-evenbly,mera-review,hauru}). 

As stressed in the review \cite{kadanoff}, the 'old' block-spin renormalization procedures and the 'new' tensor-network approaches 
have remained different in their aims : the goal of block-spin renormalization is to produce some explicit tractable RG flow
 for the few parameters of the initial Hamiltonian,
while the Tensor Network activity is more oriented towards the production of very powerful numerical algorithms,
based on the variational optimization of the whole Tensor Network representing the ground state wave function,
and where the numerical precision can be systematically improved by increasing the dimension of the elementary tensors.
Nevertheless, it seems useful to establish some bridges between the two points of view, at least on specific examples.
In the present paper, we thus consider the critical line of the XXZ antiferromagnetic chain, for which 
Martin-Delgado and Sierra have proposed a simple block-spin procedure that takes into account
the $SU_q(2)$ symmetry via the introduction of boundary fields in the intra-block Hamiltonian
\cite{delgado,jafari,langari}. Here, we describe how the obtained solution can be reformulated as the simplest  
scale-invariant Tree-Tensor-State compatible with the quantum symmetries of the model, and analyze in detail its various properties.

The paper is organized as follows.
Section \ref{sec_model} contains a reminder on the XXZ chain and the quantum group $SU_q(2)$.
In section \ref{sec_tree}, we discuss the construction of scale-invariant Tree-Tensor-State compatible with the quantum symmetries of the model.
In section \ref{sec_scaling}, the scaling properties of local operators involving one spin or two spins are analyzed, in order to compute the ground-state energy,
the magnetizations and the staggered magnetizations. The section \ref{sec_multif} is then devoted to the Shannon-Renyi entropies that characterize 
the multifractality of the components of the wave function.
Our conclusions are summarized in section \ref{sec_conclusion}.

\section{Reminder on the XXZ chain and the quantum group $SU_q(2)$  }

\label{sec_model}

\subsection{ Reminder on the XXZ chain with anisotropy $\Delta$ : continuous line of critical points  }

The XXZ chain with coupling $J>0$ and anisotropy $\Delta$ 
\begin{eqnarray}
{\cal H}_{\{J,\Delta\}} && =  \sum_{i=1}^{N-1}  H_{i,i+1}
\nonumber \\
H_{i,i+1} && = J  \left( \sigma_i^x \sigma_{i+1}^x+ \sigma_i^y \sigma_{i+1}^y + \Delta \sigma_i^z \sigma_{i+1}^z \right)
\label{xxz}
\end{eqnarray}
is one of the most studied integrable system \cite{gaudin}.
While the ground state displays Long-Ranged Ferromagnetic order for $\Delta \leq -1 $ and
Long-Ranged Anti-Ferromagnetic order for $\Delta >1$, 
 it remains critical in the whole region $-1< \Delta \leq 1$.
With the standard parametrization in terms of the phase $\gamma \in [0, \pi[$
\begin{eqnarray}
 \Delta = \cos \gamma
\label{deltaqsmall}
\end{eqnarray}
 the spin-spin correlations decay as the following power-laws with oscillating and non-oscillating contributions \cite{giamarchi}
\begin{eqnarray}
 < \sigma^x_i \sigma^x_j >  && \propto c_1 \frac{(-1)^{i-j} }{\vert i-j \vert^{\eta_x} } +c_2 \frac{ 1 }{\vert i-j \vert^{\eta_x+\frac{1}{\eta_x}} }
\nonumber \\
 < \sigma^z_i \sigma^z_j >  && \propto c_3  \frac{(-1)^{i-j} }{\vert i-j \vert^{\eta_z} } + c_4   \frac{ 1 }{\vert i-j \vert^{2} }
\label{corre}
\end{eqnarray}
where the exponents $(\eta_x,\eta_z)$ vary continuously as a function of $ \Delta = \cos \gamma $
\begin{eqnarray}
\eta_x = 1- \frac{\gamma}{\pi} = \frac{1}{\eta_z} 
\label{eta}
\end{eqnarray}
with the special points :

(i) the isotropic Antiferromagnetic chain  $\Delta=1$ corresponding to $\gamma=0$ involves the exponents $\eta_x=1=\eta_z$.

(ii) the anisotropic Antiferromagnetic case $\Delta=\frac{1}{2}$ corresponding to $\gamma=\frac{\pi}{3}$ involves the exponents $\eta_x=\frac{2}{3}$ and $\eta_z=\frac{3}{2}$.

(iii) the free-fermionic case $\Delta=0$ corresponding to $\gamma=\frac{\pi}{2}$ involves the exponent $\eta_x=\frac{1}{2}$ 

(iv) in the region of ferromagnetic interaction $-1<\Delta<0$ corresponding to $\gamma>\frac{\pi}{2} $, the correlation $< \sigma^z_i \sigma^z_j > $ is dominated by the ferromagnetic part in $1/\vert i-j \vert^{2} $.

Note that the case $J<0$ does not require an independent study since the unitary operator $U=e^{i \frac{\pi}{2} \sum_n n \sigma_n^z}$ acts on the Hamiltonian of Eq. \ref{xxz} as \cite{gaudin}
\begin{eqnarray}
U {\cal H}_{\{J,\Delta\}} U^{-1} && = - {\cal H}_{\{J,-\Delta\} }= {\cal H}_{\{-J,-\Delta\}}
\label{uxzz}
\end{eqnarray}

\subsection{Reminder :  XXZ chain with boundary fields and the quantum group $SU_q(2)$  }

The parametrization of the anisotropy in terms of $q$ as
\begin{eqnarray}
 \Delta = \frac{q+q^{-1}}{2} 
\label{deltaq}
\end{eqnarray}
and the introduction of the following opposite boundary fields on the first and on the last spins
\begin{eqnarray}
h_1 = J \frac{q-q^{-1}}{2} = - h_N
\label{h1hn}
\end{eqnarray}
gives the Hamiltonian
\begin{eqnarray}
H_{\{J,q\}} && = J \sum_{i=1}^{N-1}   \left( \sigma_i^x \sigma_{i+1}^x+ \sigma_i^y \sigma_{i+1}^y + \frac{q+q^{-1}}{2}  \sigma_i^z \sigma_{i+1}^z \right)
+  J \frac{q-q^{-1}}{2} (\sigma_1^z- \sigma_N^z)
\label{xxzq}
\end{eqnarray}
that has the nice property to commute with the generators of the quantum group $SU_q(2)$ \cite{baxter,ps}, 
the q-deformation of $SU(2)$ corresponding to the isotropic point $q=1$.
Let us summarize at the most pedestrian level what is needed for our present purposes (see \cite{ps} for more details).
The idea is to replace the local ladder Pauli matrices $\sigma_i^{\pm}$ by the q-deformed ladder operators
\begin{eqnarray}
\Sigma^{\pm}_i \equiv  &&  q^{\frac{1}{2}(\sigma_1^z +...\sigma_{i-1}^z)} 
\  \sigma_i^{\pm} \ 
q^{-\frac{1}{2}(\sigma_{i+1}^z +...\sigma_{N}^z)} 
\label{qgeneratorsi}
\end{eqnarray}
with the commutator
\begin{eqnarray}
[ \Sigma^{+}_i , \Sigma^{-}_i ]= q^{(\sigma_1^z +...\sigma_{i-1}^z)} 
\  \sigma_i^{z} \ 
q^{-(\sigma_{i+1}^z +...\sigma_{N}^z)} 
\label{qgeneratorscommutateur}
\end{eqnarray}
while for $i \ne j$ they commute
\begin{eqnarray}
[ \Sigma^{+}_i , \Sigma^{-}_j ]= 0
\label{qgeneratorscommut}
\end{eqnarray}

The corresponding global operators for the whole chain
\begin{eqnarray}
\Sigma^z \equiv  &&  \sum_{i=1}^N \sigma_i^z
\nonumber \\
\Sigma^{\pm} \equiv  &&  \sum_{i=1}^N \Sigma_i^{\pm}
\label{qgenerators}
\end{eqnarray}
satisfy the commutation relation
\begin{eqnarray}
[ \Sigma^{+} , \Sigma^{-} ]&& = \frac{q^{\Sigma^z}-q^{-\Sigma^z}} {q-q^{-1}}
\label{rule2}
\end{eqnarray}
that reduces to the the usual $SU(2)$ commutation relation for $q \to 1$.

The commutation of the Hamiltonian of Eq. \ref{xxzq} with the generators $(\Sigma^z,\Sigma^{\pm} )$
\begin{eqnarray}
[H_{\{J,q\}}  , \Sigma^z] = 0 = [H_{\{J,q\}}  , \Sigma^{\pm}] 
\label{xxzqcomm}
\end{eqnarray}
yields that the eigenstates can be classified with the representations of $SU_q(2)$.
As long as $q$ is not a root of unity, these representations are similar to the usual case of $SU(2)$
with multiplets $\vert j,m>$, where $m=-j,..,+j$ denotes the magnetization along $z$,
and where the ladder operators $ \Sigma^{\pm} $ can be used to generate the various states of the multiplets \cite{ps},
as described below for the examples of $b=2$ and $b=3$ spins.

An important technical point is that in all computations, $q$ should be treated as a formal variable that is left unchanged by complex conjugation \cite{ps}
so that the q-deformed ladder operators of Eq. \ref{qgenerators}
are the adjoint operators of each other  \cite{ps}
\begin{eqnarray}
\Sigma_i^+ = (\Sigma_i^- )^{\dagger} 
\label{adjoint}
\end{eqnarray}
On the other hand, the critical region $-1<\Delta =\cos \gamma \leq 1$ (Eq \ref{deltaqsmall}) corresponds to the complex values 
\begin{eqnarray}
 q=e^{i \gamma}
\label{qeig}
\end{eqnarray}
where the boundary fields in Eq \ref{h1hn} are imaginary \cite{ps}
\begin{eqnarray}
 h_1= -h_N  = J \frac{q-q^{-1}}{2} = i J \sin \gamma
\label{himag}
\end{eqnarray}
so that the Hamiltonian of Eq. \ref{xxzq} is actually not self-adjoint, even if its eigenvalues remain real \cite{ps}.
In summary, all the intermediate calculations have to be performed with the formal variable $q$ left unchanged by complex conjugation,
even if at the end of the day, one wishes to apply the final results of the computations to the critical region with complex $q$ (Eq. \ref{qeig}).

\subsection{ Supplementary symmetries }

\label{sec_sup}

It is useful to introduce 
the following supplementary symmetries of the Hamiltonian $H_{\{J,q\}}$ and the $SU_q(2)$ generators $\Sigma^{\pm} ,\Sigma^z $  \cite{ps} :

(1) the Relabeling symmetry where the sites $(1,2,..,N)$ are relabeled in reverse order
by $(N,N-1,..,1)$ and where $q$ is replaced by $q^{-1}$ \cite{ps}
\begin{eqnarray}
R \equiv \{ (1,2,..,N) \to (N,N-1,..,1) ; q \to q^{-1} \}
\label{relabel}
\end{eqnarray}
satisfy $R^2=1$ and can be usedl to characterize the symmetry $r_j= 1$ or antisymmetry $r_j=-1$ of multiplets
\begin{eqnarray}
R \vert j,m> = r_j  \vert j,m>
\label{rj}
\end{eqnarray}

(2) the Flip Symmetry where all the $z$-components are flipped $\sigma_i^z \to - \sigma_i^z$ and where $q$ is replaced by $q^{-1}$ \cite{ps}
relates the states of opposite magnetizations in a given multiplet (or the same state for the zero-magnetization state $m=0$)
\begin{eqnarray}
F \vert j,m> = \epsilon_j  \vert j, - m>
\label{epsj}
\end{eqnarray}

Let us now describe explicitly how all these symmetries allows to structure the Hilbert space for a small number of spins.

\subsection{ Example with $b=2$ sites }

\label{sec_b2}

For two spins, the generators of $SU_q(2)$ reduce to
\begin{eqnarray}
\Sigma^{z} && = \sigma_1^z+ \sigma_2^z
\nonumber \\
\Sigma^{\pm} && = \sigma_1^{\pm}q^{-\frac{\sigma_2^z}{2} }+ q^{\frac{\sigma_1^z}{2} } \sigma_2^{\pm}
\label{b2}
\end{eqnarray}
The four-dimensional Hilbert space can be decomposed into 

(i) the triplet $ j=1$ containing the two ferromagnetic states $ \vert ++> $ and  $ \vert --> $, while the 
third state of zero magnetization can be obtained by applying  $\Sigma^- $ to $ \vert ++> $ or $\Sigma^+$ to $ \vert --> $ and by normalizing
\begin{eqnarray}
\vert j=1,m=1 > && = \vert ++>
\nonumber \\
\vert j=1,m=0 > && = \frac{1}{\sqrt{2 \Delta} }\left( q^{-\frac{1}{2} } \vert -+> + q^{\frac{1}{2} } \vert +-> \right)
\nonumber \\
\vert j=1,m=-1 > && = \vert -->
\label{b2triplet}
\end{eqnarray}
This triplet corresponds to $r_1=+1$ and $\epsilon_1=+1$ for the supplementary symmetries of section \ref{sec_sup}.

(ii)  the singlet $j=0$  is annihilated by the two ladder operators $\Sigma^{\pm} $
\begin{eqnarray}
\vert j=0 > && = \frac{1}{\sqrt{2 \Delta} }\left( q^{\frac{1}{2} } \vert -+> - q^{-\frac{1}{2} } \vert +-> \right)
\label{b2sing}
\end{eqnarray}
and corresponds to $r_0=-1$ and $\epsilon_0=-1$.

\subsection{ Example with $b=3$ sites }

\label{sec_b3}

For three spins, the generators of $SU_q(2)$ reduce to
\begin{eqnarray}
\Sigma^{z} && = \sigma_1^z+ \sigma_2^z+ \sigma_3^z
\nonumber \\
\Sigma^{\pm} && = \sigma_1^{\pm}q^{-\frac{\sigma_2^z + \sigma_3^z}{2} }+ q^{\frac{\sigma_1^z}{2} } \sigma_2^{\pm}
q^{-\frac{\sigma_3^z}{2} }
+ q^{\frac{\sigma_1^z+\sigma_2^z}{2} } \sigma_3^{\pm}
\label{b3}
\end{eqnarray}
The eight-dimensional Hilbert space can be decomposed into 

(i) the quadruplet $ j=3/2$ of states related by the ladder operators $ \Sigma^{\pm}$
\begin{eqnarray}
\vert j=3/2,m=3/2 > && = \vert +++>
\nonumber \\
\vert j=3/2,m=1/2 > && = \frac{1}{\sqrt{4 \Delta^2-1} }\left( q^{-1} \vert -++> + \vert +-+> + q \vert ++-> \right)
\nonumber \\
\vert j=3/2,m=-1/2 > && =\frac{1}{\sqrt{4 \Delta^2-1} }\left( q \vert +--> + \vert -+-> + q^{-1} \vert --+> \right)
\nonumber \\
\vert j=3/2,m=-3/2 > && = \vert --->
\label{quad}
\end{eqnarray}
characterized by $r_{3/2}=+1$ and $\epsilon_{3/2}=+1$.

(ii) two doublets $j=1/2$ that can be distinguish by the quantum number $r=\pm 1$ describing their behavior with respect to the 
Relabeling Symmetry of Eq. \ref{rj}

(ii-a) the doublet that is symmetric $r=+1$ with respect to $R$, and that will be denoted by the simpler notation $ \vert a^{\pm} >$ in the remainder of the paper
\begin{eqnarray}
 \vert a^+ > && \equiv \vert r=1, j=1/2,m=1/2 >  =\frac{1}{\sqrt{2(2 \Delta+1)} } \left( -q^{\frac{1}{2}} \vert -++> +\left( q^{\frac{1}{2}}+ q^{-\frac{1}{2}} \right)  \vert +-+> - q^{-\frac{1}{2} }\vert ++->\right) 
\nonumber \\
 \vert a^- > && \equiv \vert   r=1, j=1/2,m=-1/2>  =\frac{1}{\sqrt{2(2 \Delta+1)} } \left( q^{-\frac{1}{2}} \vert +--> -\left( q^{\frac{1}{2}}+ q^{-\frac{1}{2}} \right)  \vert -+-> + q^{\frac{1}{2} }\vert --+> \right) 
\label{eigena}
\end{eqnarray}
with the flip antisymmetry $\epsilon=-1$.

(ii-b) the doublet that is antisymmetric $r=-1$ with respect to the Relabeling $R$,  and that will be denoted by the simpler notation $ \vert b^{\pm} >$ in the remainder of the paper
\begin{eqnarray}
 \vert b^+ > && \equiv \vert r=-1, j=1/2,m=1/2 >  = \frac{1}{\sqrt{2(2 \Delta-1)} } \left(q^{\frac{1}{2}} \vert -++> +\left( q^{\frac{1}{2}}- q^{-\frac{1}{2}} \right)  \vert +-+> - q^{-\frac{1}{2} }\vert ++->\right)  
\label{eigenb}
\\
 \vert b^- > && \equiv \vert r=-1, j=1/2,m=-1/2 >  =\frac{1}{\sqrt{2(2 \Delta-1)} } \left( -q^{-\frac{1}{2}} \vert +--> +\left( q^{\frac{1}{2}}- q^{-\frac{1}{2}} \right)  \vert -+-> + q^{\frac{1}{2} }\vert --+> \right)  
\nonumber 
\end{eqnarray}
with the flip antisymmetry $\epsilon=-1$.

\section{ Scale-invariant Tree-Tensor-State preserving the $SU_q(2)$ symmetry  }

\label{sec_tree}

\subsection{ Coarse-graining with blocks of $b=3$ spins}

In a standard block-spin renormalization, the chain of $N$ spins $\sigma_i$ is decomposed into $\frac{N}{3}$ blocks  $j=1,..,\frac{N}{3}$
containing the three spins $(\sigma_{3j-2},\sigma_{3j-1},\sigma_{3j}) $ : out of this Hilbert space of dimension $2^3=8$,
one wishes to keep only 2 states $\vert \psi_{j}^{\pm} >$ that can be parametrized by a single renormalized spin $\tau_j$.
Within the Tensor-Network perspective, this coarse-graining step is described by
the following operators called {\it isometries } (see the reviews \cite{vidal_intro,phd-evenbly,mera-review,hauru}) 
between the eight-dimensional Hilbert space of the three initial spins $(\sigma_{3j-2},\sigma_{3j-1},\sigma_{3j})$
and the two-dimensional Hilbert space of the renormalized spin $\tau_j$
\begin{eqnarray}
w_j && = \vert \tau_j^z=+1 >< \psi_{j}^+  \vert  + \vert \tau^z_j=-1 > < \psi_{j}^-  \vert 
\nonumber \\
w_j^{\dagger} && = \vert \psi_{j}^+  ><  \tau_j^z=+1 \vert  + \vert \psi_{j}^-  > < \tau_j^z=-1  \vert 
\label{transfot}
\end{eqnarray}
where the product
\begin{eqnarray}
w_j w_j^{\dagger}  && = \vert \tau_j^z=+1 >< \tau_j^z=+1  \vert  + \vert \tau_j^z=-1 > < \tau_j^z=-1   \vert = I_{\tau_j}
\label{wwdag}
\end{eqnarray}
is the identity of the renormalized Hilbert space, while the product
\begin{eqnarray}
w_j^{\dagger}w_j && = \vert \psi_j^+ ><  \psi_j^+ \vert 
 + \vert \psi_j^-> <\psi_j^- \vert \equiv P_j
\label{wdagw}
\end{eqnarray}
corresponds to the projector $P_j$ onto the subspace spanned by the two states $\psi_j^{\pm} $ that are kept out of the initial Hilbert space of the block.

For the whole chain, the correspondence between the initial Hilbert space of size $2^N$ and the renormalized Hilbert space of size $2^{\frac{N}{3}}$ 
is describe by the global operators
\begin{eqnarray}
W= \otimes_{j=1}^{\frac{N}{3}}  w_j  
\nonumber \\
W^{\dagger}= \otimes_{j=1}^{\frac{N}{3}} w_j^{\dagger} 
\label{wtransfot}
\end{eqnarray}

\subsection{ Ascending super-operator ${\cal A}$ }

\label{subsec_a}

For the initial chain of $N$ spins $\sigma_i$, the most general operator can be expanded on the Pauli basis as
\begin{eqnarray}
 O_{\sigma}  = \sum_{\alpha_1=0,x,y,z}\sum_{\alpha_2=0,x,y,z}  ...  \sum_{\alpha_N=0,x,y,z} C_{\alpha_1,\alpha_2,...,\alpha_N}
 \sigma_1^{\alpha_1} \sigma_2^{\alpha_2} ... \sigma_N^{\alpha_N} 
\label{opaulis}
\end{eqnarray}
in terms of the $4^N$ coefficients
\begin{eqnarray}
C_{\alpha_1,\alpha_2,...,\alpha_N}= \frac{1}{2^N} Trace( O_{\sigma}  \sigma_1^{\alpha_1} \sigma_2^{\alpha_2} ... \sigma_N^{\alpha_N}  )
\label{lambda}
\end{eqnarray}

After one renormalization step, the operator $O_{\sigma}$ of Eq. \ref{opaulis}
becomes projected onto the following operator acting on the renormalized spins $\tau_j$
\begin{eqnarray}
  {\hat O}_{\tau} = 
{\cal A}[ O_{\sigma}]= W O_{\sigma}  W^{\dagger} = \sum_{\alpha_1=0,x,y,z}  ...  \sum_{\alpha_N=0,x,y,z} C_{\alpha_1,\alpha_2,...,\alpha_N}
 \prod_{j=1}^{\frac{N}{3}} w_j ( \sigma_{3j-2}^{\alpha_{3j-2}} \sigma_{3j-1}^{\alpha_{3j-1}}  \sigma_{3j}^{\alpha_{3j}}   ) w_j^{\dagger} 
\label{asc}
\end{eqnarray}
This defines the ascending super-operator ${\cal A}$ (see the reviews \cite{vidal_intro,phd-evenbly,mera-review,hauru}).

In each block $j$, the projection of an operator $o_{j}=\sigma_{3j-2}^{\alpha_{3j-2}} \sigma_{3j-1}^{\alpha_{3j-1}}  \sigma_{3j}^{\alpha_{3j}} $
of the internal spins produces the following operator for the renormalized spin $\tau_j$
\begin{eqnarray}
 w_j ( o_j ) w_j^{\dagger} =
 \frac{1+\tau_j^z}{2} <  \psi_j^+ \vert o_j \vert \psi_j^+ > + \frac{1-\tau_j^z}{2} <  \psi_j^- \vert o_j \vert \psi_j^- >
 +\tau_j^+ <  \psi_j^+ \vert o_j \vert \psi_j^- > +\tau_j^- <  \psi_j^- \vert o_j \vert \psi_j^+ >
\label{wjspinswjdagger}
\end{eqnarray}

To see how it works, it is now useful to consider the simplest examples of operators containing only a few spin operators :

(0) Zero-spin operator :  the identity $I_{\sigma}$ is projected onto the identity $I_{\tau}$ as a consequence of Eq \ref{wwdag}
\begin{eqnarray}
{\cal A}[ I_{\sigma}]= W I_{\sigma}  W^{\dagger} =  I_{\tau} 
\label{azero}
\end{eqnarray}

(1) One-spin operator $\sigma_i^{\alpha}$  :  the projection of a single spin operator 
involves only the corresponding renormalized spin operators $\tau_j^{\beta}$.

(2) Two consecutive spin operators $\sigma_i^{\alpha} \sigma_{i+1}^{\alpha'}$ : 
here depending on the position of the pair with respect to the blocks, the projection will involve either a single renormalized spin $\tau_j$,
or two consecutive renormalized spins $\tau_j$ and $ \tau_{j+1} $.

(3) Three consecutive spin operators  $\sigma_i^{\alpha} \sigma_{i+1}^{\alpha'} \sigma_{i+2}^{\alpha''}$ : 
here the projection can involve either a single renormalized spin $\tau_j$,
or two consecutive renormalized spins $\tau_j$ and $ \tau_{j+1} $, but cannot produce three consecutive renormalized spins operators.

\subsection{ Choice of the isometry $w$ preserving the $SU_q(2)$ symmetry}

\label{sec_choice}

The choice of the isometry $w$ is the point where the traditional block-spin renormalization approach and the Tensor-Network perspective differ.

In the traditional block-spin renormalization approach, the isometry $w$ of Eq. \ref{transfot} is chosen by requiring that the two kept states $\vert \psi_j^{\pm}>$ are the ground-states of some 'intra-block Hamiltonian' involving only the three initial spins $(\sigma_{3j-2},\sigma_{3j-1},\sigma_{3j})$.
The 'computational advantage' is that the diagonalization of the three spins Hamiltonian usually gives simple explicit results for $\vert \psi_j^{\pm}>$ as a function of
the couplings of the Hamiltonian. The first 'theoretical drawback' is that there is some arbitrariness in the decomposition of the Hamiltonian into
'intra-block' and 'extra-block' contributions that can lead to completely different outputs, so that the quality of the results
will strongly depend on the 'cleverness' of the choice of the intra-block Hamiltonian.
The second 'theoretical drawback' is that the choice of the ground states of 
the 'intra-block Hamiltonian' does not take at all into account the 'environment' of the neighboring blocks.

In the Tree-Tensor-Network perspective (see the reviews \cite{ver,cirac,vidal_intro,phd-evenbly,mera-review,hauru,orus} and references therein), one represents instead the whole ground-state wavefunction $\vert \psi^{GS}_N>$ as a tree-tensor-state involving
 isometries, and the choice of the isometries is based on the minimization of the total energy
\begin{eqnarray}
E^{GS}_N = <  \psi^{GS}_N \vert H \vert \psi^{GS}_N>
\label{energie}
\end{eqnarray}
The 'computational drawback' is that this global minimization can usually only be done numerically.
However, the number of parameters can be drastically reduced if one takes into account all the symmetries of the model
(see the review \cite{orus} and references therein).
In the following, we describe how the symmetries of the XXZ chain allows to constraint sufficiently the possible isometries of the smallest dimension
in order to obtain an analytical solution.

(i) Translation-invariance  : for models that are translation-invariant in the bulk, the isometries $w_j$ of Eq. \ref{transfot}
are chosen to be the same in each block $j=1,..,\frac{N}{3}$.
It is important to stress however that the decomposition into blocks breaks nevertheless the equivalence between the three spins inside each block : 
this issue will be further discussed below.

(ii) Scale-invariance : for translation-invariant critical models, the coarse-graining procedure can be iterated 
with the same isometry $w$
for the  $( \frac{\ln N}{\ln 3})$ layers representing the successive renormalization steps : the initial layer $t=0$ contains the $N$ initial spins $\sigma_i$, 
the first layer $t=1$ contains $\frac{N}{3}$ renormalized spins $\tau_j$, the layer $t=2$ contains $\frac{N}{9}$ second renormalized spins... 
up to the top layer $t= \frac{\ln N}{\ln 3}$ containing a single t-renormalized spin representing the two degenerate ground-states of the whole chain.
In summary at this stage, the ground-state wavefunction $\vert \psi^{GS}_N>$ is written as a translation-invariant scale-invariant tree-tensor state
involving the same isometry $w$ everywhere in all layers.

(iii) $SU_q(2)$ symmetry : the presence of symmetries in tensor networks states is discussed in detail in the series of papers \cite{tsym,u1,su2,global,protected},
and the present $SU_q(2)$ symmetry can be considered as a deformation of the $SU(2)$ symmetry analyzed in \cite{su2}.
However, since we focus here only on tree-tensor state involving the isometry $w$ with the smallest possible dimension  (Eq. \ref{transfot}),
we do not need the full formalism described in \cite{su2} for general tensors in the presence of the SU(2) symmetry,
but we can use the language of the initial spins $\sigma$ and the renormalized spins $\tau$ to simplify the analysis.
The idea is to introduce the $SU_q(2)$ generators for the renormalized spins $\tau$ similar to Eq. \ref{qgenerators}
\begin{eqnarray}
T^z &&= \sum_{j=1}^{\frac{N}{3}} \tau_j^z
\nonumber \\
T^{\pm} &&= \sum_{j=1}^{\frac{N}{3}} T_j^{\pm}
\nonumber \\
T^{\pm}_j &&\equiv    q^{\frac{1}{2}(\tau_1^z +...\tau_{j-1}^z)} 
\  \tau_j^{\pm} \ 
q^{-\frac{1}{2}(\tau_{j+1}^z +...\tau_{\frac{N}{3}}^z)} 
\label{Tqgenerators}
\end{eqnarray}
and to require that the generators of the $SU_q(2)$ symmetry are preserved by the coarse-graining procedure as described by the ascending 
super-operator ${\cal A}$ introduced in Eq. \ref{asc}
\begin{eqnarray}
 {\cal A} (\Sigma^z) && = T^z 
\nonumber \\
 {\cal A} (\Sigma^{\pm} ) && =T^{\pm}
\label{ascqgenerators}
\end{eqnarray}
 At the level of each block $j$, 
one thus needs that the renormalized spin operators $(\tau_j^z,\tau_j^{\pm})$ are the images via the ascending superoperator ${\cal A}$
\begin{eqnarray}
\tau_j^z = {\cal A} ( \Sigma^{z}_{3j-2,3j-1,3j} ) 
\nonumber \\
\tau_j^{\pm} = {\cal A} ( \Sigma^{\pm}_{3j-2,3j-1,3j} )
\label{tauiz}
\end{eqnarray}
of the generators associated to the three initial spins of a given block
\begin{eqnarray}
\Sigma^{z}_{3j-2,3j-1,3j} && \equiv \sigma^z_{3j-2}+\sigma^z_{3j-1}+\sigma^z_{3j}
\nonumber \\
 \Sigma^{\pm}_{3j-2,3j-1,3j} && \equiv \sigma^{\pm}_{3i-2}q^{-\frac{(\sigma_{3i-1}^z +\sigma_{3i}^z)}{2}} 
+q^{\frac{\sigma_{3i-2}^z }{2}} \sigma^{\pm}_{3i-1}q^{-\frac{\sigma_{3i}^z}{2}} 
+q^{\frac{(\sigma_{3i-2}^z +\sigma_{3i-1}^z)}{2}} \sigma^{\pm}_{3i} )= \tau_j^{\pm}
\label{tauizbloc}
\end{eqnarray}

Since the renormalized spin $\tau_j$ is a doublet $j=1/2$, one obtains that the two states $\vert \psi^{\pm}>$  that are kept in each block
should also correspond to a doublet $j=1/2$. Since the $SU_q(2)$ representations for $b=2$ spins (section \ref{sec_b2}) 
only contains a triplet $j=1$ and a singlet $j=0$, one obtains that the coarse-graining with blocks of $b=2$ sites are not compatible with the $SU_q(2)$ symmetry.
Since the $SU_q(2)$ representations for $b=3$ spins (section \ref{sec_b3}) 
 contains a quadruplet $j=3/2$ and two doublets $j=1/2$ that have different quantum number $r=\pm 1$ for the relabeling symmetry,
one concludes that there are only two possible isometries that are compatible with the $SU_q(2)$ and the relabeling symmetry :

(a) the isometry $w_a$ based on the choice $\vert  \psi^{\pm}>= \vert a^{\pm}>$ of Eq \ref{eigena}
\begin{eqnarray}
w_a && = \vert \tau^z=+1 >< a^+  \vert  + \vert \tau^z=-1 > < a^-  \vert 
\label{wa}
\end{eqnarray}

(b) the isometry $w_b$ based on the choice $\vert  \psi^{\pm}>= \vert b^{\pm}>$ of Eq \ref{eigenb}
\begin{eqnarray}
w_b && = \vert \tau^z=+1 >< b^+  \vert  + \vert \tau^z=-1 > < b^-  \vert 
\label{wb}
\end{eqnarray}

So one needs to compare the ground state energies (Eq. \ref{energie}) for the two tree-tensor-states $\vert \psi^{GS}_a>$ and $\vert \psi^{GS}_b> $
based on the isometries $w_a$ and $w_b$ respectively.
Physically, one expects that this coarse-graining based on $SU_q(2)$ representations with $j=1/2$ up to the whole ground state
has a meaning only in the antiferromagnetic critical region
\begin{eqnarray}
0<\Delta \leq 1
\label{afregion}
\end{eqnarray}
while the ferromagnetic critical region $-1<\Delta<0$ would require to consider higher representations and thus isometries of higher dimensions.

\section{ Magnetic and energetic properties of the tree-tensor-state  }

\label{sec_scaling}

In this section, we describe the magnetic and energetic properties of the tree-tensor-state based on the isometry $w_a$ of Eq. \ref{wa},
while the comparison with the  isometry $w_b$ of Eq. \ref{wb} is postponed to section \ref{sec_compareb}.

\subsection{ Renormalization rules for spin operators inside a given block }

The action of the ascending super-operator ${\cal A} $ on any observable localized on a single block 
can be obtained by Eq. \ref{wjspinswjdagger}.
Since all block are the same, let us simplify the notations and write the renormalization rules for the three initial spins $(\sigma_1,\sigma_2,\sigma_3)$
in terms of the corresponding renormalized spin $\tau$ of the block.

Inside a given block of three sites, the renormalization rules are different for the three spins $(\sigma_1,\sigma_2,\sigma_3)$.
The two boundary spins operators are renormalized according to
\begin{eqnarray}
 {\cal A}(\sigma^{\pm}_1)  && =\zeta \tau^{\pm}
\nonumber \\
{\cal A}(\sigma^{\pm}_3)  && =\zeta \tau^{\pm}
\label{asboundarypm}
\end{eqnarray}
and
\begin{eqnarray}
 {\cal A}(\sigma^{z}_1)   && = \zeta \tau^z - \omega
\nonumber \\
 {\cal A}(\sigma^{z}_3) && = \zeta \tau^z + \omega
\label{asboundaryz}
\end{eqnarray}
with the notations
\begin{eqnarray}
\zeta && \equiv \frac{1+\Delta}{1+2 \Delta} 
\nonumber \\
\omega && \equiv \frac{ q- q^{-1} }{2(1+2 \Delta) }
\label{zetaom}
\end{eqnarray}
The central spin operators follow the renormalization rules
\begin{eqnarray}
 {\cal A}(\sigma^{\pm}_2)   && =\zeta_2 \tau^{\pm}
\nonumber \\
 {\cal A}(\sigma^{z}_2)  && = \mu_2 \tau^z 
\label{projcentral}
\end{eqnarray}
with
\begin{eqnarray}
\zeta_2 && \equiv - \frac{\Delta}{1+2 \Delta} 
\nonumber \\
\mu_2 && \equiv - \frac{ 1  }{1+2 \Delta } 
\label{zetaom2}
\end{eqnarray}
The consequences for the magnetizations and staggered magnetizations will be described below in section \ref{sec_magn}.

Let us now consider the operators involving two consecutive spin-operators that appear in the Hamiltonian.
The two interaction terms inside each block are renormalized into
\begin{eqnarray}
{\cal A}(\sigma^z_1\sigma^z_2)  = - \zeta - \omega \tau^z
\nonumber \\
{\cal A}(\sigma^z_2\sigma^z_3)  = - \zeta+ \omega \tau^z
\label{xzzinside}
\end{eqnarray}
while the two hoppings are renormalized into
\begin{eqnarray}
{\cal A}(\sigma^+_1\sigma^-_2  +\sigma^-_1\sigma^+_2 )  && = - \zeta - \omega \tau^z
\nonumber \\
{\cal A}(\sigma^+_2\sigma^-_3 +\sigma^-_2\sigma^+_3 )  && =  - \zeta+ \omega \tau^z
\label{hoppinginside}
\end{eqnarray}
so that the Hamiltonian $({\cal H}_{12}+{\cal H}_{23})$ associated to the two internal links of the block is renormalized into the constant
\begin{eqnarray}
e_{in} \equiv {\cal A}( {\cal H}_{12}+{\cal H}_{23} )  && =-2 J (2 + \Delta) \zeta =  - 2 J  \frac{(2 + \Delta)(1+\Delta)}{1+2 \Delta} 
\label{ahint}
\end{eqnarray}

\subsection{ Scaling dimensions of the magnetization and the staggered magnetization  }

\label{sec_magn}

For $\alpha=x,y,z$, we are interested into the scaling of the magnetizations for the full chain
\begin{eqnarray}
 {\cal M}_N^{\alpha} = \sum_{j=1}^N \sigma_j^{\alpha}
\label{magne}
\end{eqnarray}
and of the staggered magnetizations adapted to the N\'eel antiferromagnetic order 
\begin{eqnarray}
 {\cal N}_N^{\alpha} = \sum_{j=1}^N (-1)^j \sigma_j^{\alpha}
\label{neel}
\end{eqnarray}

From Eqs \ref{asboundarypm}, \ref{asboundaryz}  and \ref{projcentral}, one obtains in each block 
\begin{eqnarray}
{\cal A}  ( \sigma_{3j-2}^z+ \sigma_{3j-1}^z +\sigma_{3j}^z  )  && = (\zeta+\mu_2+\zeta ) \tau_j^z =   \tau_j^z 
\nonumber \\
 {\cal A}   ( \sigma_{3j-2}^{\pm}+ \sigma_{3j-1}^{\pm} +\sigma_{3j}^{\pm}  )   &&  = (\zeta+\zeta_2+\zeta ) \tau_j^{\pm}= \frac{ 2 + \Delta}{ 1+2 \Delta  } \tau_j^{\pm}
\label{magnprojbloc}
\end{eqnarray}
and
\begin{eqnarray}
{\cal A}   ( \sigma_{3j-2}^z- \sigma_{3j-1}^z +\sigma_{3j}^z  )  &&  = (\zeta-\mu_2+\zeta ) \tau_j^z =  \frac{ 3+2 \Delta }{ 1+2 \Delta} \tau_j^z 
\nonumber \\
{\cal A}    ( \sigma_{3j-2}^{\pm}- \sigma_{3j-1}^{\pm} +\sigma_{3j}^{\pm}  ) &&  = (\zeta-\zeta_2+\zeta ) \tau_j^{\pm}    = \frac{ 2 +3 \Delta}{ 1+2 \Delta  } \tau_j^{\pm}
\label{neelprojbloc}
\end{eqnarray}
so that the magnetization and staggered magnetization operators follow the simple RG rules
\begin{eqnarray}
{\cal A}   \left(  \sum_{j=1}^N \sigma_j^{z} \right)  && =  \sum_{j=1}^{\frac{N}{3}} \tau_j^z 
\nonumber \\
{\cal A}  \left(  \sum_{j=1}^N \sigma_j^{\pm} \right) && =  \frac{ 2 + \Delta}{ 1+2 \Delta  } \sum_{j=1}^{\frac{N}{3}} \tau_j^{\pm}
\label{magnproj}
\end{eqnarray}
and
\begin{eqnarray}
{\cal A}   \left(  \sum_{j=1}^N (-1)^j \sigma_j^{z} \right) && =   \frac{ 3+2 \Delta }{ 1+2 \Delta} \sum_{j=1}^{\frac{N}{3}}  (-1)^j\tau_j^z 
\nonumber \\
{\cal A}  \left(  \sum_{j=1}^N (-1)^j \sigma_j^{\pm} \right)   && = \frac{ 2 +3 \Delta}{ 1+2 \Delta  }\sum_{j=1}^{\frac{N}{3}}  (-1)^j \tau_j^{\pm}
\label{neelproj}
\end{eqnarray}

The iteration upon $t=\frac{\ln N}{\ln 3}$ RG steps then yields the following scaling dimensions $\delta$ with the system-size $N$
for the magnetizations per spin
\begin{eqnarray}
< \frac{1}{N} \sum_{j=1}^N \sigma_j^{z}>_{GS}  && \propto \frac{1}{N^{\delta^F_{z} }}  \ \ { \rm with } \ \ \delta^F_{z} =1
\nonumber \\
 <  \frac{1}{N}  \sum_{j=1}^N \sigma_j^{x,y}>_{GS} && =   \frac{1}{N} \left( \frac{ 2 + \Delta}{ 1+2 \Delta  } \right)^{\frac{\ln N}{\ln 3}} 
=   \frac{1}{N^{\delta^F_{x} } } 
\ \ { \rm with } \ \ \delta^F_{x} =1 - \frac{\ln \left( \frac{ 2 + \Delta}{ 1+2 \Delta  } \right)}{\ln 3}
\label{magnrg}
\end{eqnarray}
and for the staggered magnetizations per spin
\begin{eqnarray}
< \frac{1}{N}   \sum_{j=1}^N (-1)^j \sigma_j^{z} >_{GS}&& =  \frac{1}{N} \left(  \frac{ 3+2 \Delta }{ 1+2 \Delta} \right)^{\frac{\ln N}{\ln 3}} 
=   \frac{1}{N^{\delta^{AF}_{z} } } 
\ \ { \rm with } \ \ \delta^{AF}_{z} =1 - \frac{\ln \left(  \frac{ 3+2 \Delta }{ 1+2 \Delta} \right)}{\ln 3}
\nonumber \\
 < \frac{1}{N} \sum_{j=1}^N (-1)^j \sigma_j^{x,y} >_{GS} && =  \frac{1}{N} \left( \frac{ 2 +3 \Delta}{ 1+2 \Delta  } \right)^{\frac{\ln N}{\ln 3}} 
=   \frac{1}{N^{\delta^{AF}_{x}   } } 
\ \ { \rm with } \ \ \delta^{AF}_{x} =1 - \frac{\ln \left( \frac{ 2 +3 \Delta}{ 1+2 \Delta  } \right)}{\ln 3}
\label{neelrg}
\end{eqnarray}
These scaling dimensions can be compared to the exact exponents of the correlation function of Eq. \ref{corre}
corresponding to 
\begin{eqnarray}
 \delta^{Fexact}_{z}  && =1
\nonumber \\
  \delta^{Fexact}_{x} && =\frac{\eta_x+ \frac{1}{\eta_x} }{2} =  \frac{1- \frac{ArcCos(\Delta) }{\pi} }{2}  +  \frac{1 }{2 \left( 1- \frac{ArcCos(\Delta) }{\pi} \right) } 
\nonumber \\
 \delta^{AFexact}_{z} && =\frac{\eta_z}{2}= \frac{1 }{2 \left( 1- \frac{ArcCos(\Delta) }{\pi} \right) } 
\nonumber \\
 \delta^{AFexact}_{x} && =\frac{\eta_x}{2} = \frac{1- \frac{ArcCos(\Delta) }{\pi} }{2} 
\label{scaldim}
\end{eqnarray}

The numerical comparison between these scaling dimensions is presented for the 
  the isotropic point $\Delta=1$ and for the anisotropic point $\Delta=1/2$ in Tables I and II respectively.

\subsection{ Renormalization rule for the Hamiltonian }

We have already seen that the contributions of the two links inside each block $j$ are renormalized according to (Eq \ref{ahint} )
\begin{eqnarray}
{\cal A}( {\cal H}_{3j-2,3j-1}+{\cal H}_{3j-1,3j} )  && = e_{in}   
\label{ahinterne}
\end{eqnarray}
The contribution of the link between two blocks involves the independent renormalization of the boundary spin operators in each block
(Eqs \ref{asboundarypm} and \ref{asboundaryz})
\begin{eqnarray}
{\cal A}( {\cal H}_{3j,3j+1} )  && = 2 J \zeta^2 (\tau^+_j\tau^-_{j+1}  +\tau^-_j\tau^+_{j+1}) + J \Delta ( \omega  + \zeta \tau^z_j ) ( -\omega  + \zeta \tau^z_{j+1} ) 
\nonumber \\
&& =  J^R \left[  2 (\tau^+_j\tau^-_{j+1}  +\tau^-_j\tau^+_{j+1}) +  \Delta  \tau^z_j \tau^z_{j+1}  \right]
-  J \Delta \omega  \zeta  \tau^z_j + 
+ J \Delta  \omega   \zeta \tau^z_{j+1}  
+e_{ex}
\label{h2locksb}
\end{eqnarray}
with the renormalized coupling
\begin{eqnarray}
J^R && =  J \zeta^2
\label{jr}
\end{eqnarray}
and the constant contribution
\begin{eqnarray}
e_{ex} && =- J \Delta  \omega^2    =   J  \frac{ \Delta ( 1 - \Delta^2  ) }{  (2 \Delta+1)^2 }
\label{ex}
\end{eqnarray}
while the anisotropy is invariant $\Delta^R=\Delta$.

The contributions of the two boundary fields in Eq. \ref{xxzq} are renormalized into
\begin{eqnarray}
{\cal A} (  J \frac{q-q^{-1}}{2} (\sigma_1^z- \sigma_N^z) ) && =  e_{surf} +  J \frac{q-q^{-1}}{2} \zeta (\tau_1^z- \tau_{\frac{N}{3}}^z) 
\label{rgbfield}
\end{eqnarray}
with
\begin{eqnarray}
e_{surf} && =  2 J \frac{1-\Delta^2}{ 1+2 \Delta}
\label{esurf}
\end{eqnarray}

Putting everything together, one finally obtains that the whole Hamiltonian of Eq. \ref{xxzq} for a chain of $N=3 \left(\frac{N}{3}\right)$ spins
is renormalized into the same Hamiltonian for the $ \frac{N}{3}$ renormalized spins with the renormalized coupling $J^R$
up to constant contributions
\begin{eqnarray}
{\cal A} ( H^{(N)}_{\{J,q\}} ) && = \frac{N}{3} e_{in} + \left( \frac{N}{3} -1 \right) e_{ex} +e_{surf} 
 +  H^{(\frac{N}{3})}_{\{J^R,q\}} 
\label{rgh}
\end{eqnarray}

\begin{table}
\begin{center}
\begin{tabular}{|c|c|c|c|}
  \hline
  Isotropic case $\Delta=1$ & $\delta^F_{z} =\delta^F_{x}$  & $\delta^{AF}_{z} =\delta^{AF}_{x}$ & $\frac{e_0}{J} $\\
  \hline
  Tree-Tensor-State & 1 & $1 - \frac{\ln \left( \frac{ 5}{ 3  } \right)}{\ln 3} $= 0.535 & $\ \ \ \ \ - \frac{36}{23}$ = -1.565  \\
  \hline
 Exact & 1 & $ \ \ \ \ \ \ \ \frac{1}{2} $=0.5 & $1-4 \ln 2$ = -1.773 \\
  \hline
\end{tabular}
\end{center}
\caption{ Isotropic case $\Delta=1$ : the magnetization scaling dimensions $\delta^{F,AF}_{z,x} $ of the Tree-Tensor-State (Eqs \ref{magnrg} and \ref{neelrg}) are compared to the excat values of Eq. \ref{scaldim} , while the last column $ \frac{e_0}{J}$
corresponds to the ground-state-energy per monomer of Eq. \ref{egsa}.
}
\end{table}

\begin{table}
\begin{center}
\begin{tabular}{|c|c|c|c|c|c|}
  \hline
  Anisotropic case $\Delta=\frac{1}{2}$ & $\delta^F_{z} $ & $\delta^F_{x}$  & $\delta^{AF}_{z} $ & $\delta^{AF}_{x}$ & $\frac{e_0}{J} $\\
  \hline
  Tree-Tensor-State & 1 & $1 - \frac{\ln \left( \frac{ 5}{ 4  } \right)}{\ln 3} $= 0.797 & $1 - \frac{\ln 2}{\ln 3} $= 0.369  
& $1 - \frac{\ln \left( \frac{ 7}{ 4  } \right)}{\ln 3} $= 0.491 & $-\frac{3}{2}$   \\
 \hline
  Exact & 1 & $\ \ \ \ \ \ \frac{13}{12}=1.083$ & $\ \ \ \ \ \ \frac{3}{4}=0.75$   &$ \ \ \ \ \ \ \ \ \ \frac{1}{3} $=0.333 & $-\frac{3}{2}$ \\
  \hline
\end{tabular}
\end{center}
\caption{ Ansotropic case  $\Delta=\frac{1}{2}$: the magnetization scaling dimensions $\delta^{F,AF}_{z,x} $ of the Tree-Tensor-State (Eqs \ref{magnrg} and \ref{neelrg}) are compared to the excat values of Eq. \ref{scaldim}, while the last column $ \frac{e_0}{J}$
corresponds to the ground-state-energy per monomer of Eq. \ref{egsa}.
}
\end{table}

\subsection{ Ground-state-energy of the tree-tensor state }

The renormalization rule of Eq. \ref{rgh} for the Hamiltonian yields the following recursion
 for the ground-state energy $E^{GS(N)}_J = <  \psi^{GS}_N \vert H \vert \psi^{GS}_N>$ per site of the tree tensor state 
\begin{eqnarray}
\frac{E^{GS(N)}_J  }{N} = \frac{ e_{in} +e_{ex} }{3  } + \frac{1}{N} \left(  e_{surf} -e_{ex} \right) + \frac{1}{3} \left( \frac{E^{GS((\frac{N}{3}))}_{(J^R=J \zeta^2)}  }{\frac{N}{3} }  \right)
\label{rge}
\end{eqnarray}
The iteration up to the last renormalization step $t=\frac{\ln N}{\ln 3}$
\begin{eqnarray}
\frac{E^{GS(N)}_J  }{N} && = \frac{ e_{in} +e_{ex} }{3  } \left[ 1+ \frac{\xi^2}{3} + \left( \frac{\xi^2}{3}\right)^2+...  \left( \frac{\xi^2}{3}\right)^{t-1} \right]+
 \frac{1}{N} \left( e_{surf} -e_{ex}  \right) \left[ 1+ \xi^2+ \left( \xi^2\right)^2+... \left( \xi^2\right)^{t-1}  \right]
\nonumber \\
&& =  \frac{ e_{in} +e_{ex} }{3  } 
 \left[   \frac{1- \left( \frac{\xi^2}{3}\right)^{ \frac{\ln N}{\ln 3} }  }{1- \frac{\xi^2}{3}} \right]
 + \frac{1}{N} \left( e_{surf} -e_{ex} \right) \left[  \frac{1-\left( \xi^2\right)^{\frac{\ln N}{\ln 3}}}{ 1- \xi^2}  \right]
\nonumber \\
&& =  e_0 + \frac{e_1}{N} - \frac{(e_0+e_1) }{N^{1-\frac{ \ln \zeta^2 }{\ln 3}} }
\label{rgeiter}
\end{eqnarray}
yields the ground state energy per site in the thermodynamic limit $N \to +\infty$
\begin{eqnarray}
e_{0} \equiv \lim_{N \to +\infty} \left( \frac{<  \psi^{GS}_N \vert H \vert \psi^{GS}_N>}{N}  \right) =   \frac{ e_{in} +e_{ex} }{3- \xi^2  } 
 =  -  J  \frac{  (\Delta+1)^2 (5 \Delta+4)  }
{ (11 \Delta^2+10 \Delta+2)    }  
\label{egsa}
\end{eqnarray}
and the coefficient of the leading correction in $1/N$ 
\begin{eqnarray}
e_1 && = \left( \frac{ e_{surf} -e_{ex} }{ 1-\xi^2   } \right)  = J \frac{ (1-\Delta^2) }{\Delta }
\label{e1}
\end{eqnarray}

In particular for the case $\Delta=\frac{1}{2}$, the values \cite{delgado}
\begin{eqnarray}
e_{0}\left(\Delta=\frac{1}{2} \right)  && =  - \frac{3}{2}  J  
\nonumber \\
e_{1}\left(\Delta=\frac{1}{2} \right)  && =  + \frac{3}{2}  J  
\label{e0e1}
\end{eqnarray}
happens to coincide with the exact values derived in \cite{baxter}.
 Since the ground state energy per site $e_0$ in the thermodynamic limit $N \to +\infty$ is independent of the boundary conditions,
the same value $e_0$ has been also found for the XXZ chain with periodic or twisted boundary conditions \cite{razumov1,razumov2}.

\subsection{ Comparison with the tree-tensor-state based on the isometry $w_b$ }

\label{sec_compareb}

The properties of the tree-tensor-state based on the isometry $w_b$ of Eq. \ref{wb} can be computed similarly.
However in the region $(J>0,\Delta>0)$ that we consider, one obtains that $w_a$ is the good solution,
while $w_b$ would be the good solution in the region $J<0$ and $\Delta<0$ 
in relation with the symmetry $J \to -J$ and $\Delta \to -\Delta$ of Eq. \ref{uxzz}.

It is now interesting to return to the discussion of section \ref{sec_choice}
on the choice of the isometry $w$ :

(i) within the standard Block-spin Renormalization perspective, the choice between $w_a$ and $w_b$
is based on the comparison of the ground-state energy $E^{GS(N)}$ for blocks of size $N=3$ \cite{delgado},
that read
\begin{eqnarray}
E^{GS(N=3)}_a && =  e_{in}+e_{surf}= -2 J (1+\Delta)
\nonumber \\
E^{GS(N=3)}_b && =  2 J (1-\Delta)
\label{eintraabb}
\end{eqnarray}

(ii) within the Tree-Tensor-State perspective, the choice between the isometries $w_a$ and $w_b$
is based on the comparison of the ground-state-energy per site $e_{0}$ in the thermodynamical limit $N \to +\infty$
that contains the resummation over all renormalization steps (Eqs \ref{rgeiter} and \ref{egsa}).

\section{ Shannon-R\'enyi entropies of the tree-tensor state }

\label{sec_multif}

The groundstate wavefunction of manybody quantum systems 
have been found to be generically multifractal, with many studies
concerning the Shannon-R\'enyi entropies in pure quantum spin models
\cite{jms2009,jms2010,jms2011,moore,grassberger,atas_short,atas_long,luitz_short,luitz_o3,luitz_spectro,luitz_qmc,jms2014,alcaraz,c_renyi},
or focusing on the weight of the dominant configuration, for instance the N\'eel state for the present XXZ model \cite{razumov1,razumov2,jms2017}.
It is thus interesting in this section to analyze the multifractal properties of the tree-tensor-state.

\subsection{ Reminder on the multifractal formalism  }

The expansion of the tree-tensor-state in the $\sigma^z$ basis 
\begin{eqnarray}
\vert \psi^{GS}_N > = \sum_{S_1=\pm 1} \sum_{S_2=\pm 1} .. \sum_{S_N=\pm 1}
\psi(S_1,S_2,...,S_N) \vert S_1,S_2,..,S_N> 
\label{exppsiising}
\end{eqnarray}
involves the $2^N$ coefficients $\psi(S_1,S_2,...,S_N)  $.
The statistics of their weights $\vert \psi(S_1,S_2,...,S_N)   \vert^{2}  $ normalized to unity can be analyzed via the Inverse Participation Ratios $Y_p(N) $ where $p$ is the continuous parameter 
that is usually denoted by $q$ in the field of multifractality (but here the notation $q$ is already being used to represent the parameter of $SU_q(2)$)
\begin{eqnarray}
Y_p(N) \equiv   \sum_{S_1=\pm 1} \sum_{S_2=\pm 1} .. \sum_{S_N=\pm 1} \vert \psi(S_1,S_2,...,S_N)   \vert^{2p} 
\label{IPR}
\end{eqnarray}
The leading extensive behavior of the corresponding R\'enyi entropies define the generalized fractal dimensions $0 \leq D_p \leq 1$
 \begin{eqnarray}
S_p(N) \equiv  \frac{ \ln Y_p(N) }{1-p} \oppropto_{N \to +\infty} D_p (N \ln 2)
\label{renyi}
\end{eqnarray}
For $p=0$, $Y_0(N) $ simply measures the size $2^N$ of the Hilbert space leading to
\begin{eqnarray}
D_0=1
\label{d0}
\end{eqnarray}
For $p \to 1$, $Y_1(N)=1$ as a consequence of the normalization of the wave-function,
and Eq. \ref{renyi} corresponds to the Shannon entropy that defines the 'information dimension' $D_1$
 \begin{eqnarray}
S_1(N) \equiv -  \sum_{S_1=\pm 1} \sum_{S_2=\pm 1} .. \sum_{S_N=\pm 1} 
 \vert  \psi(S_1,S_2,...,S_N)    \vert^{2} \ln \vert   \psi(S_1,S_2,...,S_N)   \vert^{2}
\oppropto_{N \to +\infty} D_1 (N \ln 2)
\label{shannon}
\end{eqnarray}
For $p \to +\infty$, the sum of Eq. \ref{IPR} is dominated by the contribution of the maximal weight $\vert \psi_{max}\vert^2$,
so that the Renyi entropy directly characterizes the scaling of the maximal coefficient $\psi_{max}$ with respect to the system size
\begin{eqnarray}
S_{\infty}(N) \equiv  - \ln \vert \psi_{max} \vert^2  \oppropto_{N \to +\infty} D_{\infty} (N \ln 2)
\label{IPRpinfty}
\end{eqnarray}

\subsection{ Renormalization rule for the Renyi entropies  }

Block-spin renormalization has already been used to analyze the Renyi entropies of the pure and random quantum Ising chains \cite{c_renyi}.
Here we apply the same strategy to the Tree-Tensor-State for the XXZ chain.

The weight of a given configuration $(S_1,..,S_N)$ of the initial spins in the $(\sigma^z)$ basis
is directly related to the weight of the corresponding renormalized configuration
 $(\tau_1,...,\tau_{\frac{N}{3}})$ in the $\tau^z$ basis with the fixed values $\tau_j=-S_{3j-2}S_{3j-1}S_{3j} $
\begin{eqnarray}
 \vert \psi(S_1,S_2,...,S_N)   \vert^{2}  = \vert \psi^R(\tau_1,...,\tau_{\frac{N}{3}}  )   \vert^{2} 
 \prod_{j=1}^{\frac{N}{3}} \left[ \delta_{\tau_j=-S_{3j-2}S_{3j-1}S_{3j}} \vert <S_{3j-2} S_{3j-1} S_{3j} \vert a^{\tau_j}  \vert^{2} \right]
\label{weightrg}
\end{eqnarray}
One then obtains that
the IPR of Eq. \ref{IPR} follows the simple renormalization rule
\begin{eqnarray}
Y_p(N) =  (y_p)^{\frac{N}{3}} Y_p \left( \frac{N}{3} \right)
\label{IPRrg}
\end{eqnarray}
where $Y_p \left( \frac{N}{3} \right) $ is the IPR associated to the $\frac{N}{3}$  renormalized spins $\tau_j$,
while $y_p$ represents the IPR within a block of three spins that can be computed from Eq. \ref{eigena}
\begin{eqnarray}
y_p \equiv   \sum_{S_1=\pm 1} \sum_{S_2=\pm 1} \sum_{S_3=\pm 1} \vert <S_1 S_2 S_3 \vert a^{ \tau= -S_1 S_2 S_3}  \vert^{2p} 
= \frac{2^p (1+  \Delta)^p + 2 \cos (p \arccos(\Delta) ) }{ 2^p (1+ 2 \Delta)^p} 
\label{IPRbox}
\end{eqnarray}

The multiplicative renormalization rule of Eq. \ref{IPRrg} 
translates into the following additive renormalization rule
 for the Shannon-R\'enyi entropies of Eq. \ref{renyi}
\begin{eqnarray}
S_p(N) = \frac{\ln Y_p(N) }{1-p} 
=N  \frac{ \ln y_p }{3 (1-p) }  + S_p\left(\frac{N}{3}\right)
\label{rglogIPR}
\end{eqnarray}
The iteration upon the renormalization steps yields the extensive contribution
\begin{eqnarray}
\frac{ S_p(N) }{N} \opsimeq_{N \to +\infty}   \frac{ \ln y_p }{3 (1-p) } \left[ 1+ \frac{1}{3}+\frac{1}{3^2} + .. \right] =   \frac{ \ln y_p }{ 2 (1-p) } 
\label{rglogIPRthermo}
\end{eqnarray}
so that the final result for the generalized dimensions of Eq. \ref{renyi} reads
\begin{eqnarray}
 D_p (\Delta) =  \frac{ \ln y_p }{ 2 \ln 2  (1-p) } =  \frac{ 1 }{ 2 \ln 2  (1-p) } \ln \left(  \frac{2^p (1+  \Delta)^p + 2 \cos (p \arccos(\Delta) ) }{ 2^p (1+ 2 \Delta)^p}  \right)
\label{dpres}
\end{eqnarray}

Here a word of caution is in order for small values of $p$, since the generalized dimension of Eq. \ref{dpres} for $p=0$
\begin{eqnarray}
 D^{Tree}_{p=0} =    \frac{ \ln 3 }{ 2 \ln 2  }  \simeq 0.7925...
\label{dzero}
\end{eqnarray}
is not equal to unity as it should (Eq. \ref{d0}), because the Tree-Tensor-State does not span the whole Hilbert space,
since all strings of three identical consecutive spins $(+++)$ and $(---)$ have been projected out in the renormalization procedure.
So the Tree-Tensor-State cannot describe the region of small $p$ where the 'true' generalized dimensions 
$D_p^{true} $ are in the region
 $1 \geq D_p^{true} \geq D^{Tree}_{p=0}=\frac{ \ln 3 }{ 2 \ln 2  } $, while one may hope that it can give correct approximations in the region
of higher $p$ where the 'true' generalized dimensions $D_p^{true}(\Delta) $ are smaller than $D^{Tree}_{p=0} $
\begin{eqnarray}
D_p^{true}(\Delta)  \leq D^{Tree}_{p=0} =    \frac{ \ln 3 }{ 2 \ln 2  }  \simeq 0.7925...
\label{validity}
\end{eqnarray}

Let us thus describe the generalized dimensions of Eq. \ref{dpres} for special values of $p$ starting from infinity.

\subsection{ Dimension $D_{p=\infty}(\Delta)$ that measures the weight of the N\'eel state }

The limit $p \to +\infty$ of Eq. \ref{dpres}
\begin{eqnarray}
 D_{\infty} (\Delta) =   \frac{ 1 }{ 2 \ln 2   } \ln \left(  \frac{ 1+ 2 \Delta } {1+  \Delta  } \right)
\label{dpresinfty}
\end{eqnarray}
 characterizes the scaling of the maximal coefficient $\psi_{max}$ (Eq. \ref{IPRpinfty}), which occurs for the N\'eel state in the present antiferromagnetic XXZ chain
\begin{eqnarray}
 \vert \psi_{max} \vert^2  =  \vert \psi_{Neel} \vert^2 \oppropto_{N \to +\infty} e^{-D_{\infty} (N \ln 2) }
\label{maxneel}
\end{eqnarray}

For instance for the isotropic case $\Delta=1$ and the anisotropic case $\Delta=1/2$, Eq. \ref{dpresinfty} corresponds to the values
\begin{eqnarray}
 D_{\infty} (\Delta=1) &&=   \frac{ \ln 3 }{ 2 \ln 2   } - \frac{1}{2}  \simeq 0.2925...
\nonumber \\
 D_{\infty} (\Delta=1/2) && =   1- \frac{ \ln 3 }{ 2 \ln 2   }   \simeq 0.2075...
\label{dpresinftyex}
\end{eqnarray}

\subsection{ Correlation dimension $D_{p=2}$  }

For $p=2$, Eq. \ref{dpres} turns out to reduce to the simple fraction $1/2$ independently of $\Delta$
\begin{eqnarray}
 D_{p=2} (\Delta) =   \frac{ 1 }{ 2    } 
\label{dpres2}
\end{eqnarray}
that translates into the sum rule
\begin{eqnarray}
Y_{p=2}(N) \equiv   \sum_{S_1=\pm 1} \sum_{S_2=\pm 1} .. \sum_{S_N=\pm 1} \vert \psi(S_1,S_2,...,S_N)   \vert^{4} \oppropto_{N \to +\infty} \sqrt{2^{-N} }
\label{IPRp2}
\end{eqnarray}
It is not clear to us whether this property has a simple physical explanation.

\subsection{ Information dimension $D_{p=1}$  }

For $p=1$, the information dimension 
\begin{eqnarray}
 D_1 (\Delta)   =  \frac{ (1+2 \Delta) \ln (1+2 \Delta)-(1+ \Delta) \ln (1+ \Delta) +\Delta \ln 2 + \sqrt{1-\Delta^2} \arccos(\Delta)}{ 2 \ln 2  (1+2 \Delta) } 
\label{dpres1}
\end{eqnarray}
gives the following values for the isotropic case $\Delta=1$ and the anisotropic case $\Delta=1/2$
\begin{eqnarray}
 D_{1} (\Delta=1) &&=   \frac{ \ln 3 }{ 2 \ln 2   } - \frac{1}{6} \simeq 0.6258...
\nonumber \\
 D_{1} (\Delta=1/2) &&  = 1 - \frac{3 \ln 3 - \frac{\pi}{\sqrt{3}} }{8 \ln 2}\simeq 0.7327...
\label{dpre1ex}
\end{eqnarray}

\subsection{ Dimension $D_{p=1/2}$  }

For the value $p=1/2$, the value of Eq. \ref{dpres}
\begin{eqnarray}
 D_{p=1/2} (\Delta) = 1-   \frac{ 1 }{ 2 \ln 2   } \ln \left(  \frac{ 1+ 2 \Delta } {1+  \Delta  } \right)
\label{dpresdemi}
\end{eqnarray}
is found to satisfy the simple relation with $D_{\infty}(\Delta)$ of Eq. \ref{dpresinfty}
\begin{eqnarray}
 D_{p=1/2} (\Delta) = 1-   D_{\infty}(\Delta)
\label{demiinfty}
\end{eqnarray}
This relation is indeed expected for the isotropic model $\Delta=1$ (see the review \cite{luitz_spectro}) 
as a consequence of the expansion of the maximal configuration in the $\sigma^z$ basis into the $\sigma^x$ basis (or vice versa)
leading to the following relation between Renyi entropies in the two different basis \cite{luitz_spectro}
\begin{eqnarray}
 S^z_{p=\infty} (\Delta) = N \ln 2 - S^x_{p=1/2} (\Delta) 
\nonumber \\
 S^x_{p=\infty} (\Delta) = N \ln 2 - S^z_{p=1/2} (\Delta) 
\label{dualzx}
\end{eqnarray}
and then the $SU(2)$ symmetry for the isotropic case $\Delta=1$ yields that 
the Renyi entropies should be the same in the two basis $S^x_{p} (\Delta=1) =S^z_{p} (\Delta=1)  $ leading to Eq. \ref{demiinfty} for $\Delta=1$.
For $\Delta \ne 1$, Eq. \ref{demiinfty} can be explained for the Tree-Tensor-State as discussed below in section \ref{sec_xbasis}.

For the isotropic case $\Delta=1$ and the anisotropic case $\Delta=1/2$, Eq. \ref{dpresdemi} corresponds to the values
\begin{eqnarray}
 D_{p=1/2} (\Delta=1) &&= \frac{3}{2}   - \frac{ \ln 3 }{ 2 \ln 2   }  \simeq 0.7075...
\nonumber \\
 D_{p=1/2} (\Delta=1/2) && =    \frac{ \ln 3 }{ 2 \ln 2   }   \simeq 0.7925...
\label{dpresdemiex}
\end{eqnarray}
This last value turns out to coincide with the exact value $D_{p=1/2} (\Delta=1/2) $ computed for odd chains with periodic boundary conditions \cite{razumov1,atas_short,atas_long}. So despite the fact that the Tree-Tensor-State cannot describe the generalized dimensions in the region of small $p$ near $p=0$ (see
the discussion before Eq. \ref{validity}), we see on this specific example $\Delta=1/2$ that the Tree-Tensor-State is able to reproduce the correct
dimension for the not-so-big value $p=\frac{1}{2}$.

\begin{table}
\begin{center}
\begin{tabular}{|c|c|c|c|c|}
  \hline
 Multifractal dimensions $ D_p(\Delta)$ & $D_{\infty}$ & $D_2$  & $D_1$ & $D_{\frac{1}{2}}$ \\
  \hline
  Isotropic case $\Delta=1$ &  $\frac{ \ln 3 }{ 2 \ln 2   } - \frac{1}{2}  $ = 0.2925 &    $\frac{1}{2}$  & $\ \ \ \ \  \frac{ \ln 3 }{ 2 \ln 2   } - \frac{1}{6} $= 0.6258  & $\frac{3}{2}   - \frac{ \ln 3 }{ 2 \ln 2   } $ = 0.7075  \\
 \hline
  Anisotropic case $\Delta=\frac{1}{2}$  & $1- \frac{ \ln 3 }{ 2 \ln 2   }  $ = 0.2075& $\frac{1}{2}$    & $ 1 - \frac{3 \ln 3 - \frac{\pi}{\sqrt{3}} }{8 \ln 2}$ =0.7327   & $ \ \ \ \ \  \frac{ \ln 3 }{ 2 \ln 2   } $ = 0.7925 \\
  \hline
\end{tabular}
\end{center}
\caption{ Comparison of the multifractal dimensions $D_p(\Delta) $ of the Tree-Tensor-State for the special values $p=\infty,2,1,1/2$ between the  isotropic case $\Delta=1$ and the anisotropic case $\Delta=\frac{1}{2}$.
}
\end{table}

\subsection{ Properties of the Tree-Tensor-State in the $\sigma^x$ basis  }

\label{sec_xbasis}

Up to now, we have only considered the $\sigma^z$ basis, but in this final section it is actually interesting to discuss some properties 
of the Tree-Tensor-State in the $\sigma^x$ basis.

The two states $ \vert \tau^z=\pm 1 > = \vert a^{\pm}>$ defining the renormalized spin $\tau$ for a block of three initial spins 
that was written in the $\sigma^z$ basis in Eq. \ref{eigena} 
translates 
into the following expansions in the $\sigma^x$ basis $(\sigma_1^x=\pm 1,\sigma_2^x=\pm 1,\sigma_3^x=\pm 1)$
\begin{eqnarray}
  \vert \tau^z=+1 >_{x}  =\frac{1}{ 2\sqrt{(2 \Delta+1)} } 
( && q^{\frac{1}{2}} \vert -++> -\left( q^{\frac{1}{2}}+ q^{-\frac{1}{2}} \right)  \vert +-+> + q^{-\frac{1}{2} }\vert ++->
\nonumber \\
&& - q^{\frac{1}{2}} \vert +--> +\left( q^{\frac{1}{2}}+ q^{-\frac{1}{2}} \right)  \vert -+-> - q^{-\frac{1}{2} }\vert --+>) 
\nonumber \\
\vert \tau^z=-1 >_{x}  =\frac{1}{2\sqrt{(2 \Delta+1)} } 
( &&  q^{-\frac{1}{2}} \vert -++> -\left( q^{\frac{1}{2}}+ q^{-\frac{1}{2}} \right)  \vert +-+> + q^{\frac{1}{2} }\vert ++->
\nonumber \\
&& +q^{-\frac{1}{2}} \vert +--> -\left( q^{\frac{1}{2}}+ q^{-\frac{1}{2}} \right)  \vert -+-> + q^{\frac{1}{2} }\vert --+>) 
\label{eigenax}
\end{eqnarray}
where the two ferromagnetic states $(\sigma_1^x,\sigma_2^x,\sigma_3^x)=(+++)$ and $(\sigma_1^x,\sigma_2^x,\sigma_3^x)=(---)$
are again absent, while the other amplitudes are very reminiscent of Eq. \ref{eigena}.

Let us now write the two states of the renormalized spin $\tau$ in the $\tau^x$ basis
\begin{eqnarray}
  \vert \tau^x=+1 >_{x}  && =\frac{\vert \tau^z=+1 >_{x}+\vert \tau^z=-1 >_{x}  }{ \sqrt{2} } 
\nonumber \\
=
\frac{1}{ 2\sqrt{2(2 \Delta+1)} } 
( && \left( q^{\frac{1}{2}}+ q^{-\frac{1}{2}} \right)  \vert -++> -2 \left( q^{\frac{1}{2}}+ q^{-\frac{1}{2}} \right)  \vert +-+> + \left( q^{\frac{1}{2}}+ q^{-\frac{1}{2}} \right) \vert ++->
\nonumber \\
&& - \left( q^{\frac{1}{2}}- q^{-\frac{1}{2}} \right)  \vert +--> +0  +\left( q^{\frac{1}{2}}- q^{-\frac{1}{2}} \right)\vert --+>) 
\nonumber \\
\vert \tau^x=-1 >_{x} && = \frac{\vert \tau^z=+1 >_{x} -\vert \tau^z=-1 >_{x}  }{ \sqrt{2} } 
\nonumber \\
  =\frac{1}{2\sqrt{2(2 \Delta+1)} } 
( &&  \left( q^{\frac{1}{2}}- q^{-\frac{1}{2}} \right) \vert -++> + 0  - \left( q^{\frac{1}{2}}- q^{-\frac{1}{2}} \right)\vert ++->
\nonumber \\
&& -\left( q^{\frac{1}{2}}+ q^{-\frac{1}{2}} \right) \vert +--> +2 \left( q^{\frac{1}{2}}+ q^{-\frac{1}{2}} \right)  \vert -+-> - \left( q^{\frac{1}{2}}+ q^{-\frac{1}{2}} \right)\vert --+>) 
\label{eigenaxx}
\end{eqnarray}
For the isotropic case $\Delta=1=q$, these states are similar to the expressions in the $z$-basis with only three components of the corresponding magnetization (Eq. \ref{eigena}), while for the anisotropic cases $\Delta \ne 1$, they are clearly different.
As a consequence the Renyi entropies $S^z_p(\Delta) $ and $S^x_p(\Delta)$ in the two basis are not expected to be the same for 
the anisotropic cases $\Delta \ne 1$ and generic index $p$. 
However in the limit $p \to +\infty$, where only the weight of the maximal component $\psi_{max}$ corresponding to the N\'eel state in the two basis
remains (Eq. \ref{maxneel}), some simplification occurs in the $\sigma^x$ basis : 
the N\'eel state $\vert +-+>_x$ for the block only occurs in the renormalized state $ \vert \tau^x=+1 >_{x}  $,
while the opposite N\'eel state $\vert -+->_x$ for the block only occurs in the renormalized state $ \vert \tau^x=-1 >_{x}  $,
and both weights are equal to the corresponding weight in the $z$-basis (Eq \ref{eigena})
\begin{eqnarray}
\vert _x< +-+ \vert \tau^x=+1 >_{x} \vert^2 = \vert _x< -+- \vert \tau^x=-1 >_{x} \vert^2 = \frac{1+\Delta}{1+2 \Delta} = 
\vert _z< +-+ \vert \tau^z=+1 >_{x} \vert^2 
\label{neelxz}
\end{eqnarray}
These properties lead to the equality of the Renyi entropies $S^z_p(\Delta) $ and $S^x_p(\Delta)$ for $p \to +\infty$
\begin{eqnarray}
 S^z_{p=\infty} (\Delta) = S^x_{p=\infty} (\Delta) 
\label{zxpinfty}
\end{eqnarray}
and explain the simple relation found in Eq. \ref{demiinfty} for the Tree-Tensor-State.

\section{ Conclusions  }

\label{sec_conclusion}

For the line of critical antiferromagnetic XXZ chains with coupling $J>0$ and anisotropy $0<\Delta \leq 1$,
we have obtained that the translation-invariant scale-invariant Tree-Tensor-State of the smallest dimension
compatible with the quantum symmetries of the model corresponds to the block-spin renormalization procedure
 introduced previously by Martin-Delgado and Sierra 
 via the introduction of boundary fields in the intra-block Hamiltonian to make it $SU_q(2)$ invariant \cite{delgado,jafari,langari}.
Even if the final output is the same, we have explained the differences in reasoning between the two points of view :
in the block-spin renormalization approach, the solution is based on the lowest eigenstate $E^{GS(N=3)}_a $ for the block of $N=3$ spins (Eq. \ref{eintraabb}),
while in the Tree-Tensor approach, the chosen solution is based on the best ground-state-energy per site $e_0$ in the thermodynamic limit $N \to +\infty$,
after the resummation of the contributions of all the renormalization scales (Eqs \ref{rgeiter} and \ref{egsa}).
As a consequence, the two criteria could turn out to be different in other models.
In the future, it would be thus interesting to see for other critical quantum spin chains whether the Tree-Tensor approach agrees also with some previous choices of block-spin renormalization or produces new solutions.

In the second part of the paper, we have described the energetic and magnetic properties of the Tree-Tensor-State, as well as the 
as the multifractality of the components of the wave function. Despite its simplicity, the Tree-Tensor-State turns out to be able to reproduce
some exact results for the anisotropic case $\Delta=1/2$, like the ground-state energy (Eq. \ref{e0e1}) or the multifractal dimension 
$D_{p=1/2} $ of Eq. \ref{dpresdemiex}. Further work is needed to understand whether 'disentanglers' between blocks preserving the $SU_q(2)$
symmetry can be introduced besides the block-coarse-graining in order to produce some simple analytical MERA for the XXZ chain.

\end{document}